\documentclass[preprint]{elsarticle}
\usepackage{ifpdf}
\usepackage[usenames]{color}
\usepackage{amssymb}
\usepackage{amsmath}
\usepackage{subfigure}
\usepackage{graphicx}
\usepackage{epstopdf}
\usepackage{subfigure}
\usepackage{color}
\begin{document}


\begin{frontmatter}

\title{Observability of Market Daily Volatility}

\cortext[cor2]{Corresponding author.}

\author[UC]{Filippo Petroni\corref{cor2}}
\ead{fpetroni@unica.it}

\author[UAQ]{Maurizio Serva}

\address[UC]{Dipartimento di Scienze Economiche ed Aziendali,
Universit\`{a} di Cagliari, Italy}

\address[UAQ]{Dipartimento di Ingegneria e Scienze 
dell'Informazione e Matematica, Universit\`a 
dell'Aquila,  Italy}

\begin{abstract}
We study the price dynamics of 65 stocks from the Dow Jones Composite 
Average from 1973 until 2014. We show that it is possible to define a 
Daily Market Volatility $\sigma(t)$ which is directly observable from
data. This quantity is usually indirectly defined by 
$r(t)=\sigma(t) \omega(t)$ where the $r(t)$ are the daily returns of the 
market index and the $\omega(t)$ are 
i.i.d. random variables with vanishing average and unitary variance.
The relation $r(t)=\sigma(t) \omega(t)$ alone is unable to give
an operative definition of the index volatility, which remains unobservable.
On the contrary, we show that using the whole information
available in the market, the index volatility can be operatively 
defined and detected.
\end{abstract}

\end{frontmatter}

\newpage
It is well know that stock market returns 
are uncorrelated on lags larger than a single day. This is an 
inescapable consequence of the efficiency of markets.
On the contrary, absolute returns have memory for very
long times, a phenomenon known as volatility clustering.
These phenomena are very well documented in the literature and 
known as stylized facts \cite{Gui97, DaPe12, DaPe12b, DaPe11}.
 
Furthermore, there is a large empirical evidence that volatility 
auto-correlations decay hyperbolically \cite{Ta, DGE, BB, BCdL, Pa}
and there is also a growing evidence that
volatility signals have a multi-fractal nature
\cite{PSa, PSb, PSc, BPSVV, WYHS, Mi}.

Daily historical volatility is an unobservable variable and is usually measured by the absolute value
of daily returns which are instead observable. This definition gives only an approximation of the
real volatility $\sigma(t)$ which can be indirectly defined by 
$r(t)=\sigma(t) \omega(t)$ where $r(t)$ are the daily returns of the 
market index and the $\omega(t)$ are 
i.i.d. random variables with vanishing average and unitary variance.
The relation $r(t)=\sigma(t) \omega(t)$ alone is unable to give
an operative definition of the index volatility, which remains unobservable.
We will show that using the whole information
available in the market, the index volatility can be operatively 
defined and detected, i.e., we will define an observable volatility
for a market index which exhibits all the statistical features expected for 
this variable.

The daily returns of single stock (say $\alpha$) are given by
$r_{\alpha}(t)= \ln[S_\alpha(t)/S_\alpha(t\!-\!1)]$ where $S_\alpha(t)$ 
is the closing price of stock $\alpha$ at day $t$.
Then, if one wants to extract volatility from data one can consider that
$r_{\alpha}(t)= \sigma_{\alpha}(t)\omega_{\alpha}(t)$
where the $\omega(t)$ have vanishing averages and unitary variance.
Volatility $\sigma_{\alpha}(t)$
can be eventually extracted considering the high frequency (intra-day) 
continuous trading, but the problem remains highly unsolved because 
of the overnight contribution to the return $r_{\alpha}(t)$
for which there is not continuous trading. 
Therefore, the best measure of (historical) daily volatility 
simply remain the absolute returns $|r_{\alpha}(t)|$.

If the aim is to measure the global volatility of a market,
we will show that things can be different.
One can address the problem considering volatility of
a proper representative index.
Nevertheless, if one considers a price-weighted index 
(as Nikkei 225) the main contributions 
will be artificially given by those stocks with a larger price.
The problem is circumvented if one considers a capitalization-weighted 
index (as Hang Seng) or an equally-weighted index 
(as S$\&$P 500 EWI).
The daily return $r(t)$ of this last index is simply the plain average of 
the returns of its components. i.e.

\begin{equation}
r(t)=  \frac{1}{N}
\sum_{\alpha=1}^{N} r_{\alpha}(t)
\label{ewi-return}
\end{equation}
where $N$ is the number of stocks in the basket and
$r_{\alpha}(t)= \ln(S_\alpha(t)/S_\alpha(t\!-\!1))$ and $S_\alpha(t)$ 
is the daily closing price of the stock $\alpha$ at day $t$.
For the other two types of index the difference is
that the average is weighed by price or by capitalization.

Again, the underlying index daily volatility $\sigma(t)$ is not directly 
observable from daily returns but it is indirectly defined by
$r(t)= \sigma(t)\omega(t)$. 
Because of market efficiency,
it can be assumed that the $\omega(t)$ are independent identically
distributed random variables with vanishing average
and unitary variance. 
One could argue that $\sigma(t)$ is, indeed, observable from 
high frequency data, but, as already mentioned,
the problem of the overnight contribution to the daily returns remains.

Therefore, since daily market volatility is not objectively given 
by the index return (only the product $\sigma(t)\omega(t)$ is
observable),  its distribution  depends on the model chosen for 
$\omega(t)$.
Gaussianity is often assumed
as in ARCH-GARCH modeling (the leptokurticity of the distribution 
of returns is, in this case, entirely charged to volatility).
Nevertheless, one can do other choices for the distribution of $\omega(t)$
as, for example, the uniform distribution (between $-\sqrt3$
and $\sqrt3$ in order that the variance is unitary).

In this paper we consider  $N=65$ titles of Dow Jones from 1973 to 
2014 so that $1 \le t \le T \simeq 10000$.
Dow Jones as an index is not equally weighted
but we can construct ourselves an EW Dow Jones index
for which the daily returns $r(t)$ are simply the plain average of 
returns of its components as in definition (\ref{ewi-return}).

The absolute returns are then given by
\begin{equation}
|r(t)|= \sigma(t)|\omega(t)|=
\frac{1}{N}\left| \sum_{\alpha=1}^{N} r_{\alpha}(t) \right|
\label{absolute-return}
\end{equation}
which are the absolute returns
of the associated equally-weighted index
(for price-weighted and capitalization-weighted indexes
the only difference is that some weights must be introduced).

The core of this paper is the definition of the volatility as
\begin{equation}
\sigma^{}(t)=  \frac{1}{\sqrt3 N}
\sum_{\alpha=1}^{N} |r_{\alpha}(t)|
\label{average-volatility}
\end{equation}
so that  

\begin{equation}
\omega(t)=  \frac{r(t)}{\sigma^{}(t)}
\label{omeha-uniform}
\end{equation}
where $r(t)$ and $\sigma(t)$ are defined in equation (\ref{ewi-return})
and equation (\ref{average-volatility}).

Most of the models assume that $\sigma(t)$ and $|\omega(t+\tau)|$ are 
uncorrelated for any value of the lag 
$\tau$ (negative, positive or vanishing) or they assume
(as ARCH-GARCH) a short range correlation
(a correlation only for small values of $|\tau|$.)
Moreover $|r(t)|$ and $|\omega(t+\tau))|$ should be uncorrelated 
for any non vanishing $\tau$ as 
well as $|\omega(t)|$ and $|\omega(t+\tau)|$.

\begin{figure}[!ht]
 	\centering
 	\includegraphics[width=3.8truein]{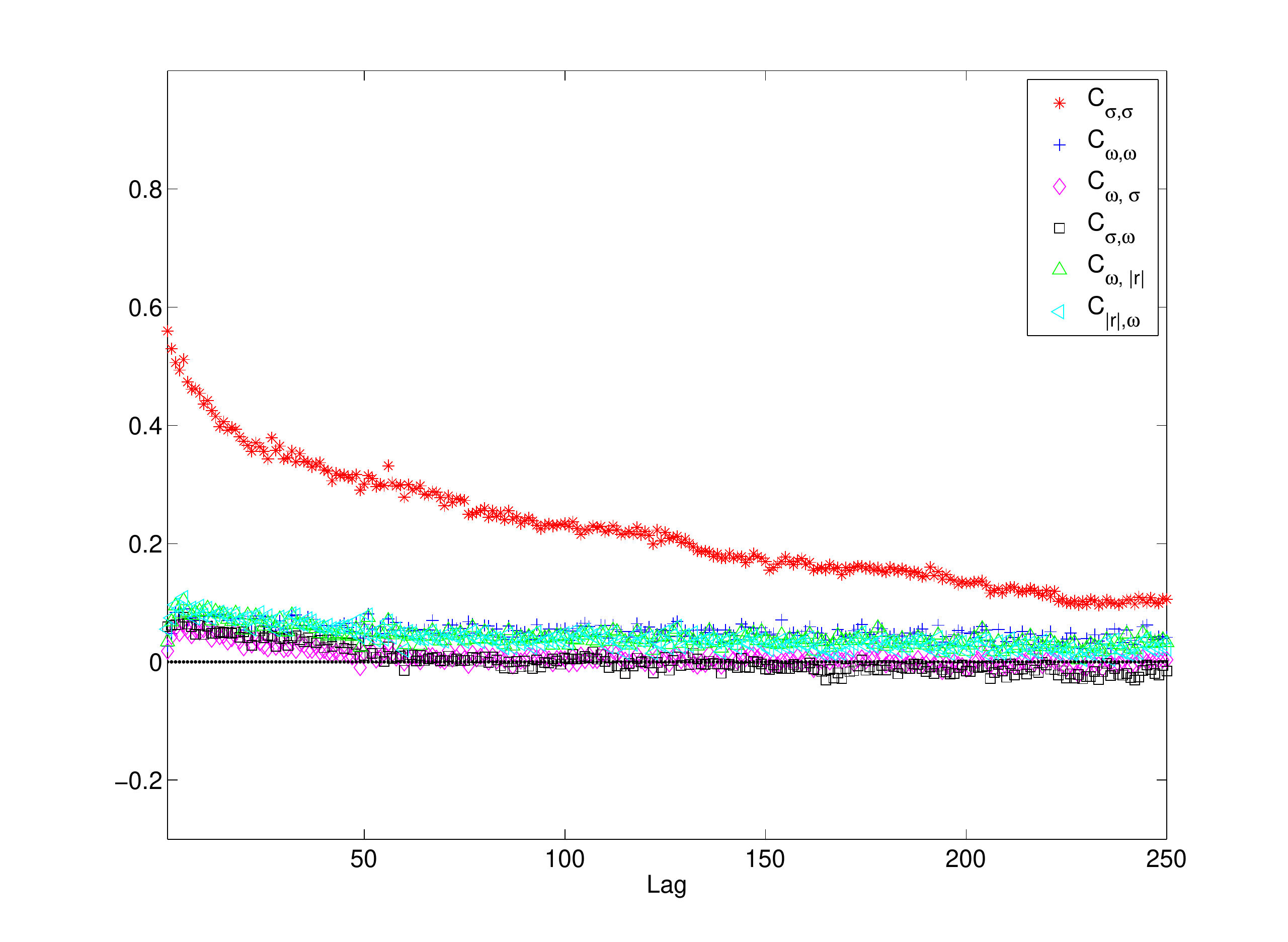}
 	\caption{Auto-correlation of $\sigma$,  $|\omega|$ and all 
the cross-correlations between $\sigma$, $\omega$ and $|r|$. 
It can be noticed that all except the auto-correlation of $\sigma$ are vanishing.}
 \label{fig1}
\end{figure}

Therefore, the first step is to show that all these
properties holds according to our definition of volatility.
After having computed the correlations 
$C_{\omega,\sigma}=C(\omega (t), \sigma (t+\tau))$,
$C_{\sigma, \omega}=C(\sigma(t), \omega(t+\tau))$
and also $C_{\omega,\omega}$, $C_{\omega,|r|}$ and $C_{|r|,\omega}$,
we have plotted them in Fig. \ref{fig1}. 
As it can be well appreciated 
the four above considered cross-correlations and the auto-correlation
substantially vanish.
This is better appreciated if compared with the volatility
auto-correlation $C_{\sigma,\sigma}$
which is also plotted and which, on the contrary, exhibits a strong lag dependence and it is significantly positive for lags up
to 250 working days.

\begin{figure}[!ht]
	\centering
	\includegraphics[width=3.8truein]
	{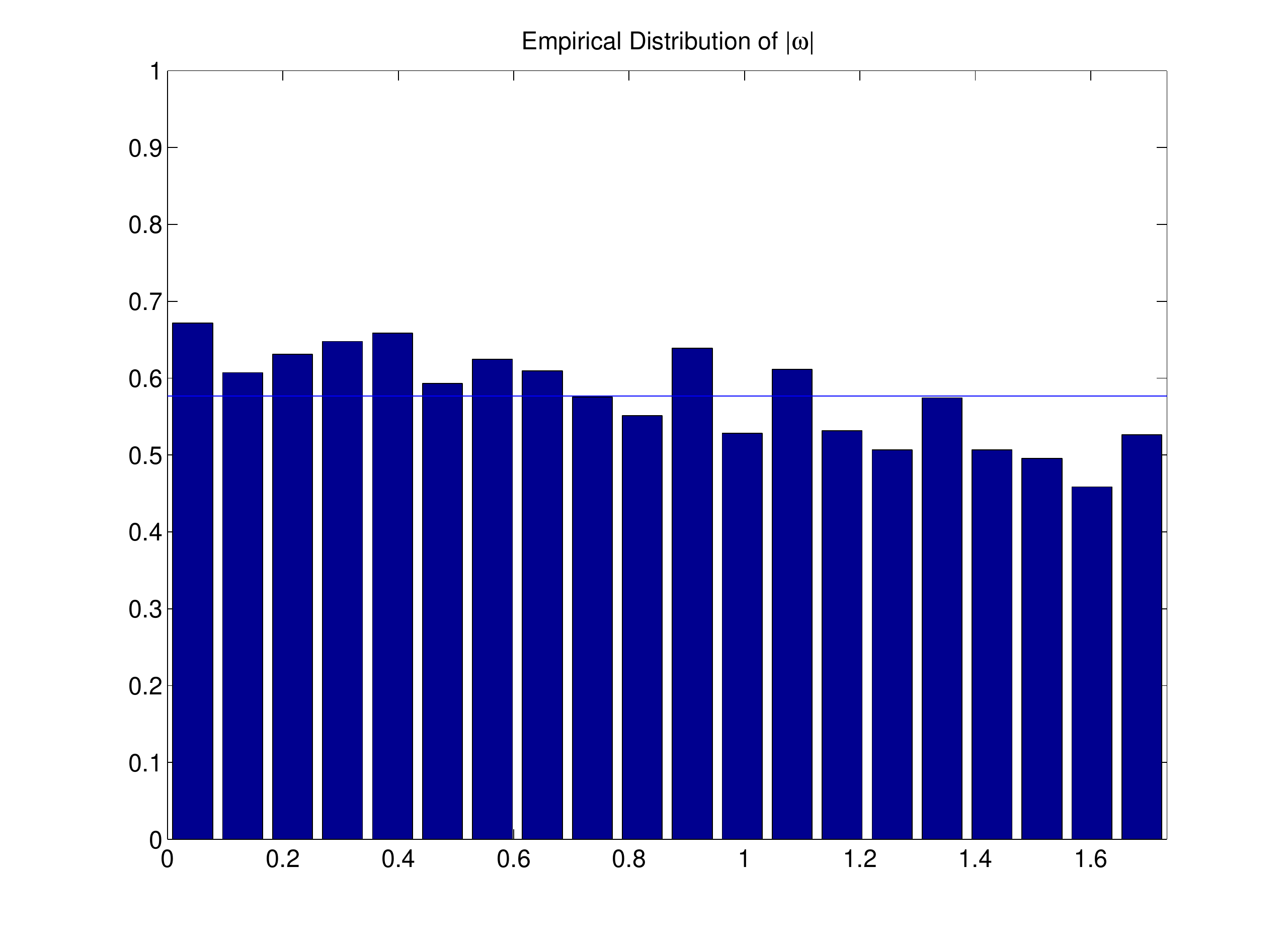}
	\caption{Uniform (almost) probability density of the $|\omega|$. 
		One has $\langle \omega^2  \rangle= 0.924 \simeq 1$) and  
		$\langle |\omega| \rangle= 0.823 \simeq \sqrt3/2 =0.866$.}
	\label{fig2}\
\end{figure}

Once shown that the variables $\omega(t)$ are independent from each other,  
from the absolute returns (2) and from the volatilities (3), 
we need to show that they have vanishing expected value 
($\langle \omega (t) \rangle=0$) and unitary variance
($\langle \omega^2 (t) \rangle=1$).
Indeed, we obtain a better result, in fact,
the distribution of the $\omega(t)$ is uniform in the range $[-\sqrt3, \sqrt3 ]$ 
(which implies vanishing expected value and unitary variance 
but also implies  $\langle |\omega (t)| \rangle=\sqrt3/2$).
This result, which is our second step, can be appreciated in Fig. \ref{fig2}
where the empirical distribution or $|\omega|$,
as defined by (4), is plotted. 


We still have a third step, to complete our argument.
Having assumed complete independence of the $\omega(t)$ 
from the volatilities one forcefully 
has that the auto-correlation $C_{|r|,|r|}$
only differs for a positive multiplicative constant $0< k < 1$
from the auto-correlation $C_{\sigma ,\sigma}$
at any time lag $\tau \ge 1$.
In fact, the auto-correlation of the absolute returns is
\begin{equation}
C_{|r|,|r|}(\tau)= \frac{\langle |r(t+\tau)||r(t)|\rangle-
\langle |r(t)|\rangle^2}
{\langle r^2(t)\rangle-\langle |r(t)|\rangle^2}
\label{abs-corr}
\end{equation}
then taking into account that $|r(t)|=\sigma(t)|\omega(t)|$
and assuming mutual independence between the $\sigma(t)$ and the $\omega(t)$,
one has for any $\tau \ge 1$:
\begin{equation}
C_{|r|,|r|}(\tau)= k \, C_{\sigma,\sigma}(\tau)
\label{corr-corr}
\end{equation}
where $C_{\sigma,\sigma}(\tau)$ is the volatility auto-correlation
\begin{equation}
C_{\sigma,\sigma}(\tau)= \frac{\langle \sigma(t+\tau)\sigma(t)\rangle-
\langle \sigma(t)\rangle^2}
{\langle \sigma^2(t)\rangle-\langle \sigma(t)\rangle^2}
\label{vol-corr}
\end{equation}
and $k$ is the constant
\begin{equation}
k= \frac{\langle \sigma^2(t)\rangle-  \langle\sigma(t)\rangle^2}
{3\langle \sigma^2(t)\rangle/4 -\langle \sigma(t)\rangle^2}
\label{kappa}
\end{equation}
where we have used the uniform distribution value
$\langle|\omega(t)|\rangle^2
/\langle\omega^2(t)\rangle=3/4$.
We compute from sample $\langle \sigma^2(t)\rangle=0.00008583$
and $\langle \sigma(t)\rangle=0.008388$ so that $k=1/2.85$.

 \begin{figure}[!ht]
 	\centering
 	\includegraphics[width=3.8truein]
 	{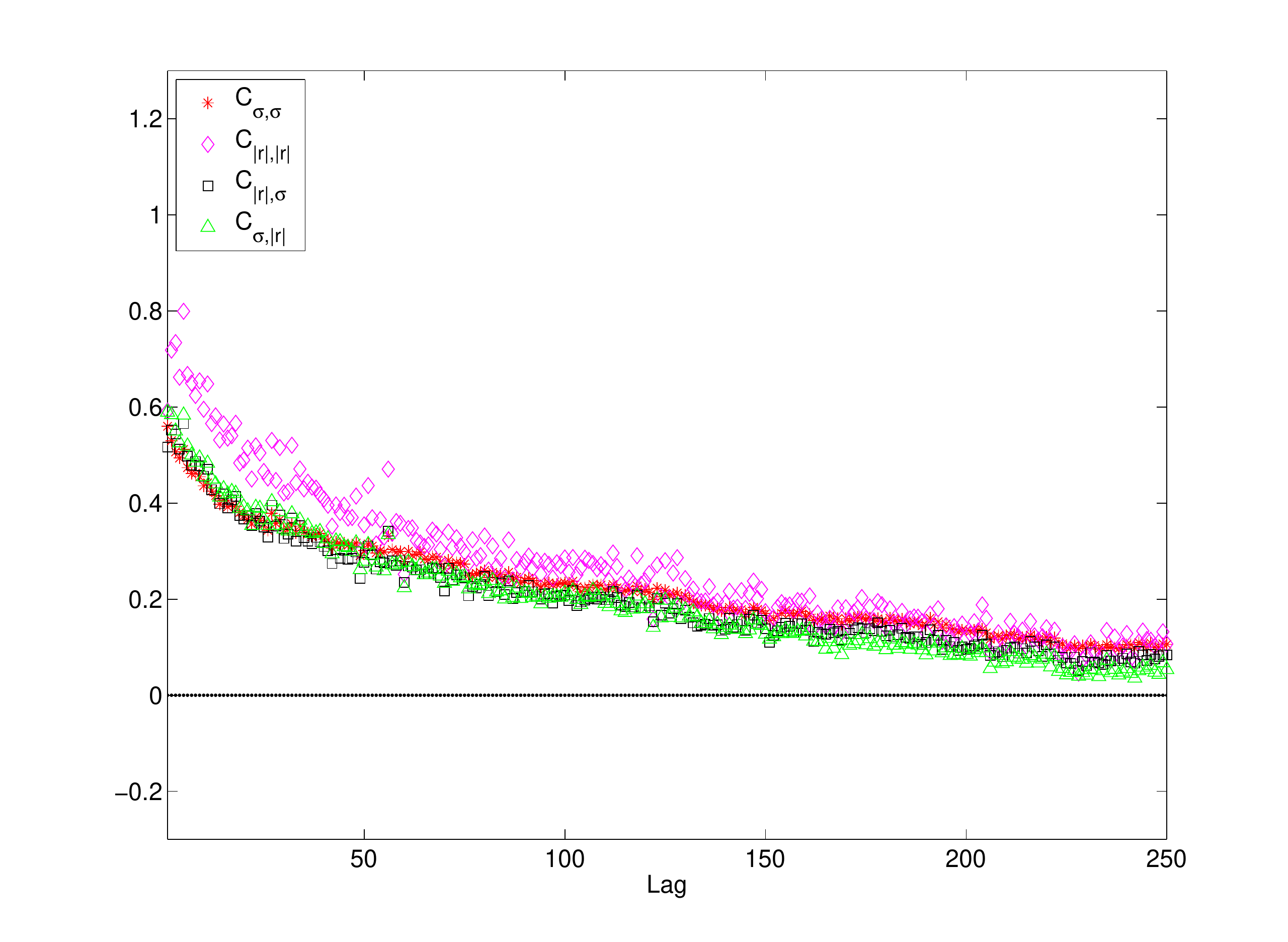}
 	\caption{Auto-correlations of $\sigma$ and $|r|$ and their 
cross-correlations. 
It is evident that the four correlations are (almost) identical once rescaled.}
 	\label{fig3}
 \end{figure}
Moreover, following the same steps, one can easily compute

\begin{equation}
C_{|r|,\sigma}(\tau)= C_{\sigma, |r|}(\tau)=
\sqrt{k} \, C_{\sigma,\sigma}(\tau)
\label{corr-corr2}
\end{equation}
where $k$ is the same value (8) previously computed ($k \simeq 1/2.85$)
so that (6) and (9) are very strict requirements. It turns out that
both relations (6) and (9) hold, in fact, 
the four correlations, rescaled by $k$ (or by $\sqrt{k}$),
are plotted in Fig. \ref{fig3} where it can be appreciated that they are 
almost identical for all values of the lag $\tau$.

Notice, that the coefficient $k$ is not a fitting parameter, but 
it is in dependently derived by market data.
The fact that after rescaling the four correlations coincide
ultimately confirms that
it is correct to write $r(t) = \sigma(t) \omega(t)$ where $\omega(t)$ and 
$\sigma(t)$ are the mutually independent variables we have defined.

In sum, we have defined the volatilities $\sigma(t)$
and, consequently, the variables $\omega(t)$
so that they (a) are mutually independent, (b) the $\omega(t)$ 
are also independent from absolute returns,
(c) the $\omega(t)$ are i.i.d.
uniformly distributed with vanishing expected value and unitary variance,
(d) the auto-correlation of the volatility
exhibits a strong lag dependence and it is significantly 
positive for lags up to 250 working days,
(e) the correct scaling of cross-correlations and auto-correlations involving
the $r(t)=\sigma(t) \omega (t)$ and the $\sigma(t)$ holds.
Therefore, we have given a definition of daily market volatility 
which is observable
and keep all the statistical features expected for this variable.

We can conclude by saying that while for a single stock it is impossible to 
extract the volatility from return using 
$r(t) = \sigma(t) \omega(t)$, we have found
a simple way to do that for an index using all absolute returns 
of the single stocks entering in the index.

\newpage


\end{document}